\documentclass[apj]{emulateapj}
\usepackage{times} 
\usepackage{color}
\usepackage{graphicx, subfigure}
\usepackage{txfonts}
\usepackage{ulem}
\usepackage{comment}
\usepackage{tabularx}
\usepackage{xcolor}
\newcommand{\msun}{\mbox{M$_\odot$}~}
\usepackage{natbib}

\newcommand{\gc}{\bf{}}
 \renewcommand{\gc}{} 

\shorttitle{Low-Mass Stars and Brown Dwarf survey in Upper~Sco}
\shortauthors{van der Plas et al.}

\begin{document}

\title{Dust masses of disks around 8 Brown Dwarfs and Very Low-Mass Stars in Upper Sco OB1 and Ophiuchus}


\author{G. van der Plas\altaffilmark{1,2}} 
 \email{info@gerritvanderplas.com} 
\author{F. M\'enard\altaffilmark{3,4}}
\author{K. Ward-Duong\altaffilmark{5}}
\author{J. Bulger\altaffilmark{5,6}}
\author{P.M Harvey\altaffilmark{7}}
\author{C. Pinte\altaffilmark{3,4}}
\author{J. Patience\altaffilmark{5}}
\author{A. Hales\altaffilmark{8,9}}
\author{S. Casassus\altaffilmark{1,2}}

\altaffiltext{1}{DAS, Universidad de Chile,, camino el observatorio 1515 Santiago, Chile}
\altaffiltext{2}{Millennium Nucleus "Protoplanetary Disks"}
\altaffiltext{3}{UMI-FCA, CNRS/INSU, France (UMI 3386), and Dept. de Astronom\'{\i}a, Universidad de Chile, Santiago, Chile.}
\altaffiltext{4}{Univ. Grenoble Alpes, IPAG, F-38000 Grenoble, France}
\altaffiltext{~}{CNRS, IPAG, F-38000 Grenoble, France}
\altaffiltext{5}{School of Earth and Space Exploration, Arizona State University, PO Box 871404, Tempe, AZ 85287, USA}
\altaffiltext{6}{Subaru Telescope, NAOJ, 650 N. A'ohoku Pl., Hilo, HI 96720, USA}
\altaffiltext{7}{Univ. of Texas at Austin, Austin, TX 78712, USA}
\altaffiltext{8}{Atacama Large Millimeter / Submillimeter Array, Joint ALMA Observatory, Alonso de C\'ordova 3107, Vitacura 763-0355, Santiago, Chile}
\altaffiltext{9}{National Radio Astronomy Observatory, 520 Edgemont Road, Charlottesville, Virginia, 22903-2475, USA}

\begin{abstract}
We present the results of ALMA band 7 observations of dust and CO gas in the disks around 7 objects with spectral types ranging between M5.5 and M7.5 in Upper Scorpius OB1, and one M3 star in Ophiuchus. We detect unresolved continuum emission in all but one source, and the $^{12}$CO J=3-2 line in two sources. We constrain the dust and gas content of these systems using a grid of models calculated with the radiative transfer code MCFOST, and find disk dust masses between 0.1 and 1 M$_\oplus$, suggesting that the stellar mass / disk mass correlation can be extrapolated for brown dwarfs with masses as low as 0.05 \msun. The one disk in Upper Sco in which we detect CO emission, 2MASS J15555600, is also the disk with warmest inner disk as traced by its H - [4.5] photometric color.  Using our radiative transfer grid, we extend the correlation between stellar luminosity and {\gc mass-averaged} disk dust temperature originally derived for stellar mass objects to the brown dwarf regime to $\langle T_{dust} \rangle \approx 22 (L_{*} /L_{\odot})^{0.16} K$, applicable to spectral types of M5 and later. This is slightly shallower than the relation for earlier spectral type objects and yields warmer low-mass disks. The two prescriptions cross at 0.27 \mbox{L$_\odot$}, corresponding to masses between  0.1 and 0.2 \msun depending on age.
\end{abstract}




\section{Introduction}

Circumstellar (and likely proto-planetary) disks are ubiquitous around solar-mass stars \citep{williams11}. It is also now well established that such disks are common around young brown dwarfs (BD) and very low mass stars (VLMS) \citep{2014A&A...570A..29B}.  As the size of the sample of such disks has grown, it is becoming increasingly possible to examine astrophysically significant trends.  For example, one important question is whether there is a relation between the disk mass and the central stellar (or sub-stellar) mass {\gc \citep{2013ApJ...773..168M, andrews2013}}.  This, of course, is directly related to the probability of planet formation around such objects as well as the likely size of young planets.  A second related question is the lifetimes of these disks, both for the dust and the gas components.  Understanding the mass and lifetimes of the gas and dust disks around a wide range of mass of the central object is key to understanding the evolution of the disks and the physical processes that produce this evolution. The most obvious of these processes include grain growth, dust settling, and photo-evaporation of the gas in addition to accretion onto the central object and onto any planets forming within the disk. The sizes and masses of BD disks may also provide clues to the details of their formation mechanism(s) as well.  For example, if sub-stellar objects form by different processes than hydrogen-burning stars such as ejection of stellar embryos \citep[e.g.][]{2003MNRAS.339..577B, 2001AJ....122..432R} or erosion of star-forming clouds by radiation from massive stars \citep{1998MNRAS.300.1205W}, then the disks may be proportionally smaller than those around higher mass objects.

The difficulty to observe disks around BDs and VLMS is linked to the most obvious difference between T Tauri-like stars (of order 1 M$_{\odot}$) and much lower mass objects, that is their luminosities.  The disks around BDs and VLMS will naturally be both fainter and colder, even if the same mass were present in both. Since both theoretical considerations {\it and} early observations have suggested some level of $M_{disk}$--$M_{star}$ relation, the observational issue is likely even more biased against easy detectability of such low-mass disks.  

Disks around BDs have up to now mostly been studied at infrared wavelengths, which trace warm dust on the disk surface close to the star and can be used to constrain the inner disk shape and lifetime \citep[e.g.][]{luhman2012}. To measure the disk mass one has to move to longer, sub-mm, wavelengths. Given the expected weak sub-mm emission from the disks around BDs, only a handful of these disks have been detected at sub-mm wavelengths to date \citep[see ][]{2006ApJ...645.1498S, 2008A&A...486..877B, andrews2013, ricci2014, 2015arXiv150904589R}. 

Recent studies using {\it Spitzer} and {\it Herschel} have investigated substantial samples of BD/VLMS \citep{2012ApJ...755...67H, 2014A&A...570A..29B, 2013A&A...559A.126A}. This paper describes the first half of an ALMA program to extend the first two of these studies to the sub-mm spectral region. This entire program was motivated by the availability of a sample of ``older'' BD/VLMS in Upper Sco from \citet{2012ApJ...755...67H} with some much younger BD/VLMS observed in the program of \citet{2014A&A...570A..29B}. Here we report on results from the ALMA observations of the $\sim$ 11 Myr old Upper Sco sample (and one Ophiuchus object) from \citet{2012ApJ...755...67H}.  Our results complement and extend the results of \citet{2014ApJ...787...42C} who report on ALMA continuum observations of K-type and early M-type stars in Upper Sco.

In \S \ref{sec:targetselection} and \S \ref{sec:observations} we describe our choice of targets and the observational details including both continuum  and $^{12}$CO J=3-2 observations. In \S \ref{sec:results} we list our basic results, whose modeling we describe in \S \ref{sec:analysis} and then discuss in \S \ref{sec:discussion}.

\section{Target selection}
\label{sec:targetselection}


Upper Sco is the youngest part of the nearest OB association Sco-Cen \citep{1999AJ....117..354D}.  The most recent age estimates are {\gc $\approx$ 10 Myr \citep{2015arXiv151008087D} and} 11 $\pm$ 2 Myr based on comparisons with the HR diagram \citep{2012ApJ...746..154P}, although ages derived from HR diagram fitting may be systematically older than activity indicators and kinematic traceback analysis \citep{2012AJ....144....8S}. While absolute ages are more difficult to determine, the relative age of Upper Sco is unambiguously older than star-forming regions such as Taurus and Ophiuchus. 
Using {\it Spitzer} data the disk population in the Upper Sco association has been estimated 
by \citet{luhman2012} at $\sim$ 500 members.
The presence of detectable circumstellar disks at this age makes this region ideal to study the late stages of disk evolution over a substantial range of stellar masses.  Indeed, based on {\it Spitzer} and {\it WISE} data, \citet{luhman2012} were able to show that disk lifetimes are likely longer at lower masses.

Our sample was selected based on the results of the far-IR {\it Herschel} PACS survey of \citet{2012ApJ...755...67H} that targeted stellar/substellar objects across multiple nearby star-forming regions. The Upper Sco sample of \citet{2012ApJ...755...67H} consists of 11 targets that possess the largest mid-IR excesses initially drawn from the results of the 35 object Upper Sco {\it Spitzer} IRS and MIPS survey of \citet{2007ApJ...660.1517S}. In order to maximize the likelihood of detecting disk emission from these late-M type targets at sub-millimeter wavelengths, our sample includes 7 of the 8 targets that were detected in both PACS channels at 70~$\mu$m and 160~$\mu$m.  Our target list also includes one object later identified as belonging to Ophiuchus, 2MASS J16214199.

The massive star population of Upper Sco has been identified with Hipparcos measurements \citep{1999AJ....117..354D} and the membership probability refined \citep{2012MNRAS.421L..97R}. The low mass population membership, however, is incomplete due to the large spatial extent of the OB association \citep[e.g.][]{2001AJ....121.1040P, 2008ApJ...688..377S}. 
The seven Upper Sco targets for this study have spectral types ranging from M5.5 to M7.5 based on spectroscopic measurements with an uncertainty of $\sim$0.5 in spectral type \citep{2004AJ....127..449M, 2000AJ....120..479A}. {\gc Our sample represents 7.2\%  of the 97 objects in the spectral type range between M4 and M8 in Upper Sco that show excess emission in the W2 and [4.5] bands as counted by \citet{luhman2012}}. Table \ref{table1} summarises the target properties including spectral type. The transition from stellar to substellar objects is expected to occur at a spectral type of M6 based on evolutionary tracks that predict a temperature of $\sim$3000 K for a 75 M$_{\mathrm{Jup}}$ object at 10 Myr \citep{1998A&A...337..403B} and the effective temperature-spectral type scale of \citet{luhman2003}. All the targets are near or below the stellar/substellar boundary. Three of the targets have been included in a proper motion survey, and two have high probabilities of membership (90\% to 100\%; 2MASS J15555600, 2MASS J16024575), while one, 2MASS J15560104, has a membership probability of only 51.4\% \citep{2009AA...504..981B}.

Another factor that may influence disk properties is the presence of a companion. One target, 2MASS J16024575, has a similar brightness companion at a separation of $\sim$0.12" \citep{2005ApJ...633..452K} {\gc \citep[18 au for a distance of 144 pc,][]{1999AJ....117..354D}} and another target, 2MASS J16193976, is a binary or possibly triple system with the B and C components at a separation of 0.18" and 2.60" {\gc (25 and 374 au)}, and a position angle of 310 and 141 degrees, respectively \citep{2006A&A...451..177B}. The remaining five targets are single objects within the detection limits compiled from several studies in \citet{2012ApJ...757..141K}.


\section{Observations and Data Reduction}
\label{sec:observations}

ALMA Early Science Cycle 1 {\gc band 7} observations were conducted on 2014 Mar 20 and 2014 Apr 27 with 2519 and 2588 seconds of total execution time respectively, translating to 4.55 minutes per source. {\gc During each execution all sources were observed, leading to comparable sensitivities and beam sizes for all targets.} The array configuration provided baselines ranging between 15.1 and 558.2 meters. During the observations the precipitable water vapor in the atmosphere was stable between 0.8 and 1.1~mm with clear sky conditions, resulting in a median system temperature of 175~K.

Three of the four spectral windows of the ALMA correlator were configured in Time Division Mode (TDM) to maximise the sensitivity for continuum observations (128 channels over 1.875~GHz usable bandwidth) in minimum observing time. The three TDM spectral windows were centred at 331.8~GHz, 333.8~GHz, and 343.8~GHz. The fourth spectral window was configured in Frequency Division Mode (FDM) to target the  $^{12}$CO J=3-2 line, using 1.875~GHz total bandwidth and an individual channel width of 488.281~kHz  (0.42 km s$^{-1}$), corresponding to a spectral resolution of  976.562 kHz (0.847 km s$^{-1}$) after Hanning smoothing.  Details of the observations and calibration are summarised in Table \ref{table2}. We estimate the absolute flux calibration, based on the scatter of flux values of the calibrators in the different executions, to be accurate within $\sim$15\%. 

\begin{table*}
\begin{center}
\caption{Observational details.\label{table2}} \smallskip
\small
\begin{minipage}[t]{\textwidth}
\noindent\begin{tabularx}{\columnwidth}{@{\extracolsep{\stretch{1}}}*{7}{l}@{}}
\hline\hline
UT Date & Number & Baseline Range & pwv &  \multicolumn{3}{c}{Calibrators} \\ 
 & Antennas & (m) & (mm) & Flux & Bandpass & Gain  \\
\hline
2014 Mar 20	& 34 & 15.1 to 437.8 & 0.97 & Titan & J1517-2422 & J1626-2951 \\
2014 Apr 27 & 35 & 22.2 to 558.2 & 1.20 & Titan & J1517-2422 & J1517-2422 \\
\hline
\end{tabularx}
\end{minipage}
\tablecomments{During data reduction we flagged 2 and 5 antennas for the March 20 and April 27 datasets, respectively.}
\end{center}
\end{table*}


We used the standard reduction tools within the \textit{Common Astronomy Software Applications} package \citep[CASA, ][]{2007ASPC..376..127M} to calibrate and combine the data. We extract the fluxes using the CASA task \textit{uvmodelfit}, while we image the disks with the CLEAN task in CASA \citep{1974AAS...15..417H} using natural weighting due to the faintness of our targets which results in a synthetic beam size of 0.56" $\times$ 0.44" at PA = -82 degrees (Figure \ref{fig:continuum}). Finally, we shift our images using the proper motion correction from the PPMXL Catalog \citep{2010AJ....139.2440R}, whose values are repeated in Table \ref{table1}.




\section{Results}
\label{sec:results}

\begin{table*}
\setlength{\tabcolsep}{3pt} 
\caption{Properties and ALMA band 7 results of sample stars. Reported uncertainties are 1 $\sigma$ values,\\ the CO uncertainty is calculated assuming a line width of 14 km s$^{-1}$. \label{table1}} \smallskip
\scriptsize
\begin{minipage}[t]{\textwidth}
\noindent\begin{tabularx}{\columnwidth}{@{\extracolsep{\stretch{1}}}*{10}{l}@{}}
\hline\hline
2MASS ID & Alternative & Sp.Type &  pmRA$^{a}$ & paDEC$^{a}$  & M$_*^{b}$  & F$_{885 \mu m}$   & F$_{CO J=3-2}$  & Multiplicity$^{c}$ & Refs. \\
       &    ID     &          & mas/yr     & mas/yr       & M$_\odot$    & mJy               & Jy km s$^{-1}$  & (au)   & \\
\hline
J15555600-2045187 & usd155556   & M6.5	&	-17.9 & -24.9    &	0.07	& 0.68 $\pm$ 0.13 & 0.275 $\pm$ 0.025 &	S	&	1; 6\\
J15560104-2338081 &	usd155601   & M6.5	&	-19.0 & -35.5	 &	0.07	& 2.46 $\pm$ 0.12 & $\leq$ 0.045 &	S	&	1; 6	\\
                  & usd155601B  &       &	-19.0 & -35.5	 &	        & 0.83 $\pm$ 0.13 & $\leq$ 0.037 &  S	&	        \\
J15591135-2338002 &	usco128	    & M7  	&	9.9   &  -13.6	 &	0.06	& 2.00 $\pm$ 0.12 & $\leq$ 0.051 &	S	&	2; 6	\\
J16024575-2304509 & usco55	    & M5.5	&	-20.8 &  -34.3	 &	0.10	& 0.49 $\pm$ 0.13 & $\leq$ 0.047 &	18	&	2; 5	\\
J16060391-2056443 &	usd160603   & M7.5	&	-11.7 &  -24.5	 &	0.04	&      $\leq$ 0.13& $\leq$ 0.049 &	S	&	1; 6	\\
J16100541-1919362 &	usd161005   & M7	&	-6.8  &  -37.7	 &	0.06	& 0.52 $\pm$ 0.13 & $\leq$ 0.041 &	S	&	1; 6	\\
J16193976-2145349 &	usd161939   & M7	&	-9.4  &  -6.0	 &	0.06	& 0.94 $\pm$ 0.12 & $\leq$ 0.037 &	25 and 374 &	1; 4	\\
J16214199-2313432 &	Allers 8    & M3	&	-11.2 &  -29.8	 &	0.34	& 3.68 $\pm$ 0.13 & 0.350 $\pm$ 0.036& 	&	3	\\  
\hline
\end{tabularx}
\end{minipage}
\tablecomments{
$^{a}$ Proper motion values from the PPMXL catalog \citep{2010AJ....139.2440R}. Typical errors on the proper motion are 5 mas / yr. $^{b}$ Stellar masses extrapolated from the 10Myr BHAC15 models \citep{2015A&A...577A..42B} using the spectral type to temperature conversion from \citet{luhman2003}.$^{c}$ S = observed single, {\gc projected separations given in au based on the distance of 144 pc from \citet{1999AJ....117..354D}.}}
\tablerefs{(1) \citet{2004AJ....127..449M}; (2) \citet{2000AJ....120..479A}; (3) \citet{2011ASPC..448..633G}; (4) \citet{2006A&A...451..177B}; (5) \citet{2005ApJ...633..452K}; (6) \citet{2012ApJ...757..141K}.}
\end{table*}

We detect 7 out of 8 observed sources in the continuum and 2 out of 8 in the $^{12}$CO J=3-2 line. We reach a 1 $\sigma$ sensitivity of 0.13 mJy/{\gc beam} for the continuum maps, and 9 mJy{\gc /beam} per 0.423 km s$^{-1}$ channel for the CO line images. {\gc The continuum sensitivities are reached by combining both the continuum and the line spectral windows in those sources where we don't detect CO emission. For the two sources with CO emission detected, we exclude those channels within 0.1 GHz of the CO line center.} One source in our sample, 2MASS J16214199, is likely to be a member of Ophiuchus instead of Upper Scorpius OB1. This source is detected both in continuum and CO. We describe the results for this source in $\S$ \ref{sec:results} and $\S$ \ref{sec:analysis}, but do not discuss it in $\S$ \ref{sec:discussion}.

\subsection{The Band 7 continuum data}

\begin{figure*}
    \epsscale{1.2}
    \plotone{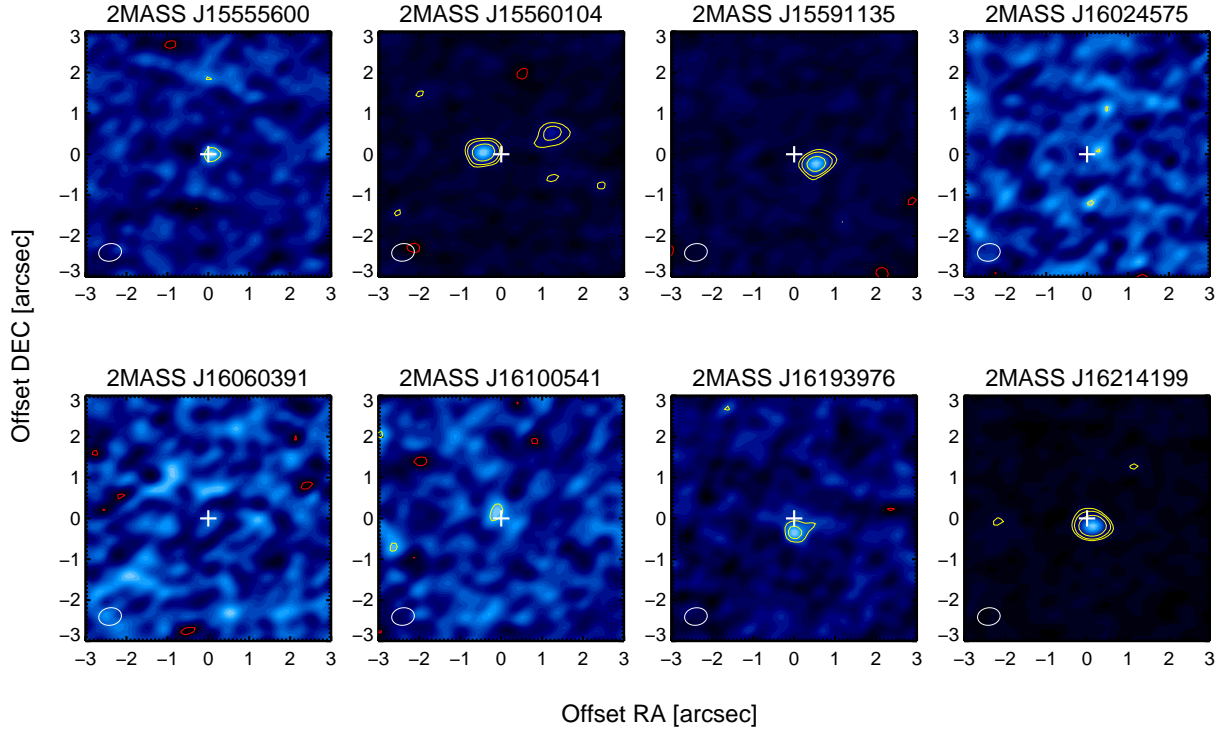}
    \caption{885 $\mu$m continuum maps of the 8 sources described in this manuscript. For each map, the pointing center [0,0] has been shifted following the proper motion values listed in Table \ref{table1} and is shown by the white plus sign. The synthesized beam size (white ellipse) is shown in the bottom left hand corner. All plots show -3 (red) and 3, 5 and 10 (yellow) $\sigma$ contours. The offset of all disks from the [0,0] coordinate is within the proper motion errors reported.}
    \label{fig:continuum}
\end{figure*}

We list the extracted flux densities in Table \ref{table1} and show 885 $\mu$m continuum maps for all targets in Figure \ref{fig:continuum}. We detect 7 sources above 3$\sigma$, as well as a 5$\sigma$ point-like signal that is offset from 2MASS J15560104 by a distance of 1.75" at position angle of 284 degrees. Visual inspection of the K=17.1 and H=17.2 magnitudes deep UKIDSS images does not show a counterpart at those wavelengths, and we consider it possible that this emission comes from a background source instead of e.g. originating from a heavily extincted (edge on) disk. We do not detect continuum emission at the location of the binary companions around 2MASS J16193976, limiting any possible disk emission from that source to be below 0.49 mJy (3 $\sigma$).  All continuum emission is spatially unresolved, limiting the detected disk emission to be within a radius of $\approx$ 40 au. We note that the continuum emission for 2MASS J16024575 is detected above 3$\sigma$ when fitting in the uv-plane, but that there are similar strength features visible in the field, depending on the weighting scheme used to reconstruct those images.


\subsection{$^{12}$CO J=3-2 line emission} 

\begin{figure*}[ht!]
    \begin{center}
    \epsscale{1.1}
    \plotone{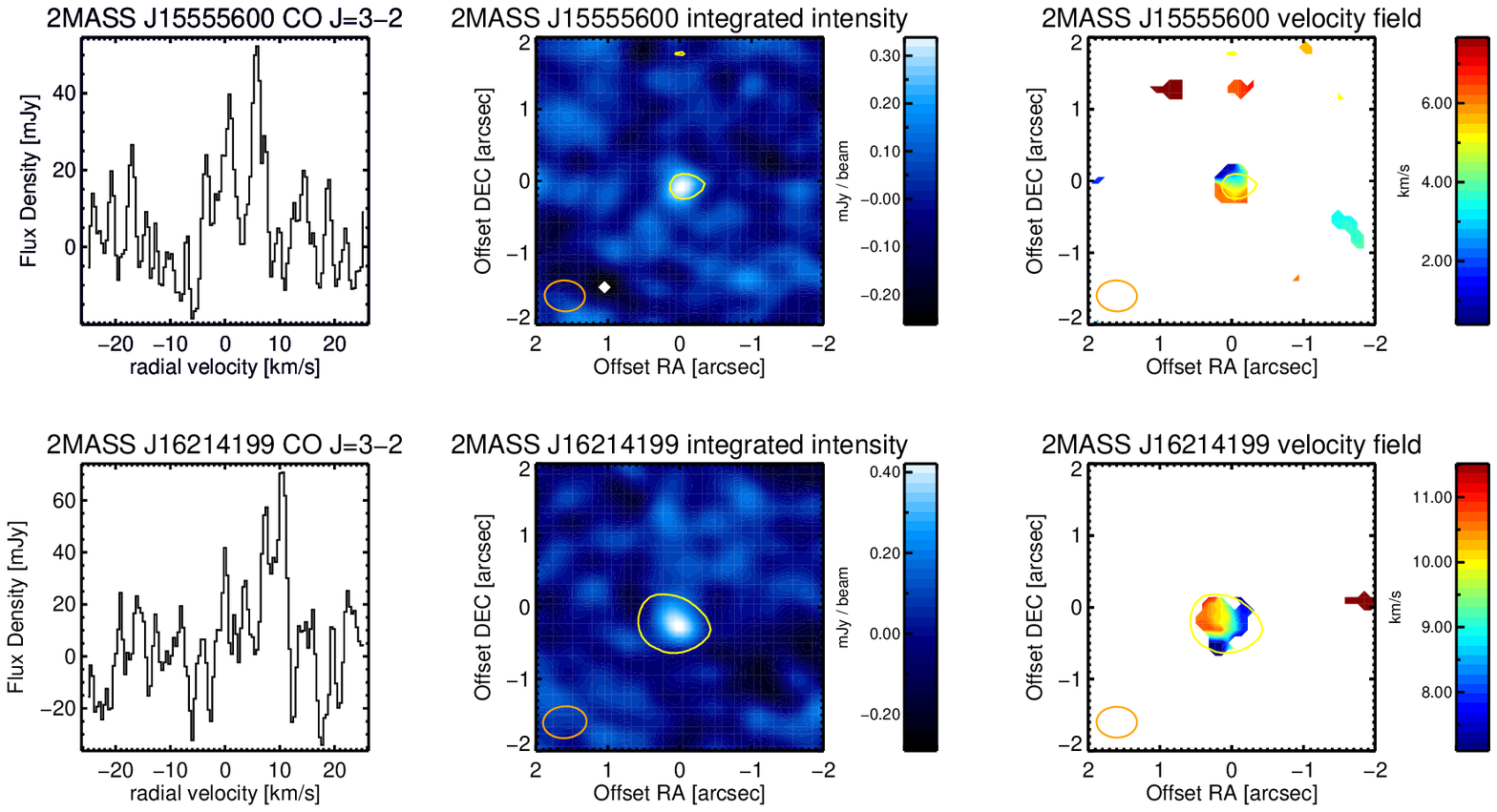}
    \caption{$^{12}$CO J=3-2 line profile (left), integrated intensity (moment 0, center) and velocity field (moment 1, right) maps for 2MASS J15555600 (top row) and 2MASS J16214199 (bottom row). The moment 0 maps have been made by collapsing all channels between -20 and 20 km s$^{-1}$, the moment 1 maps using only those parts of the maps containing emission above 3$\sigma$. All moment maps have a 3$\sigma$ continuum emission contour overplotted in yellow. The beam is shown in orange at the bottom left.}
  \label{fig:line}
  \end{center}
\end{figure*}

We detect CO emission above 3$\sigma$ intensity from the disks around 2MASS J15555600 and 2MASS J16214199. We extract the CO emission for these sources using an elliptical aperture with a major and minor axis diameter of 0.90" $\times$ 0.72" for 2MASS J15555600, and 1.35" $\times$ 0.99"  for 2MASS J16214199, respectively. The CO emission in 2MASS J15555600 and 2MASS J16214199 exceeds 3$\sigma$ levels per channel between 0.0 and 8.0 km~s$^{-1}$ and -0.5 and 11.5 km~s$^{-1}$, and the line fluxes integrated between these velocities are 175~$\pm$~20 mJy km~s$^{-1}$ and 333~$\pm$~33 mJy km~s$^{-1}$, respectively. This is a lower limit to the line flux. If we relax the constraints and integrate between -5.0 and 8.5 km~s$^{-1}$ and -2.0 and 12.5 km~s$^{-1}$, the line fluxes increase to 275~$\pm$~25 mJy km~s$^{-1}$ and 350~$\pm$~36 mJy km~s$^{-1}$, respectively. For all other disks we define a 1$\sigma$ upper limit to the CO line by using an extraction ellipse that encompasses the CO emission of both 2MASS J15555600 and 2MASS J16214199 and a linewidth of 14 km~s$^{-1}$. The velocity field (gradient) for both detections is consistent with a disk in Keplerian rotation. We show the integrated emission lines, the integrated intensity (moment 0) maps and the velocity field (moment 1) maps in Figure \ref{fig:line} and list the CO detections and 1$\sigma$ upper limits in  Table \ref{table1}.

\setcounter{footnote}{0}

\section{Radiative transfer models}
\label{sec:analysis}

The data presented above provide new constraints at long wavelengths. They allow to explore in more detail the structure of disks at the faint end of the stellar mass spectrum. In particular, we can verify and improve the dust mass estimates previously provided by \citet{2012ApJ...755...67H} based on {\it Herschel} PACS data, and compare with the dust masses estimated with analytical prescriptions (\S \ref{sec:RTvsAndrews}).

We have modeled the detected disks with the radiative transfer code MCFOST \citep{2006A&A...459..797P, 2009A&A...498..967P}, choosing a parameter space that encloses most documented cases \citet[e.g.][and references therein]{williams11}. MCFOST is a 3D, Monte-Carlo based radiative transfer code. For this paper, parametric disk models were constructed to fit the observed spectral energy distributions (SED). The approach used here is similar to the one we previously used \cite[see][]{2008A&A...486..877B, 2012ApJ...744L...1H, 2012ApJ...755...67H, 2013A&A...559A.126A}. The disks we model have low masses and are largely optically thin\footnote{{\gc See for example the 2$^{nd}$ panel of Figure 13 from \citet{2012ApJ...744L...1H}. This Figure is made for a disk twice as heavy as the most massive disk in the parameter space we explore, and shows the optical depth of the disk as function of radius for several wavelengths. For this disk at the wavelength of our observations, the disk becomes optically thick at $\approx$0.5 au. The median value for the mass of our best-fit disks is about ~20 times lower than the disk used for said Figure, justifying the assumption that our modelled disks only become optically thick at 885 $\mu$m at a fraction of an au.}}, even at MIR-FIR wavelengths \citep[see][]{2012ApJ...755...67H}. This implies that most of their geometrical parameters are not well constrained by SED modeling alone. Despite these limitations, the dust mass and to some extent the inner disk geometry that is probed most directly by the optically thick NIR part of the SED can be constrained to some degree in the modeling. 


In all our models we have fixed the dust properties. We use distributions 
of compact particles made of astronomical silicates \citep{1984ApJ...285...89D} with minimum size $a_{min}$ = 0.05$\mu$m,  maximum size $a_{max}$ = 1mm, and size distribution $dn(a)\propto a^{-3.5}da$.
The continuum opacity of this grain population is 3.72cm$^2$/g at $\lambda=885 \mu$m. 
The surface density profile is described by a power-law $\Sigma(r) \propto r^{p}$ with p in the range [-1.5, -0.5] {\gc in steps of 0.5.
The scale height profile is given by h(r) = H$_0$(r/r$_0$)$^{\beta}$ au, with r$_0$ = 100 au and H$_0$ values of [2, 5, 10, 15, 20] and $\beta$ in the range [1.0, 1.2] with 5 linear intervals.
We set the disk outer radius to 100 au in absence of strong constraints on disk size, and explore the effects of shrinking outer radii in \S \ref{sec:RTvsAndrews}.
Other parameters that we allowed to vary are the dust disk mass and inner radius. Both varied in logarithmic space in 7 equal steps between $10^{-4}$ and $10^{-7}$ M$_\odot$ for the disk mass, and for the inner radius between 5 $\times$ $10^{-3}$ and 0.25 au. The inclination angle is allowed to varied over 10 equally spaced bins in cosine (corresponding to  randomly distributed inclinations in 3D), and the stellar parameters T$_*$ and R$_*$ are restricted to values between 2800 and 3200 K and 0.2 and 2 R$_\odot$ respectively, each in 7 steps to match the temperature range expected for our spectral type range, see again \citet{2012ApJ...755...67H} and also \citet{luhman2012}. Finally, we allow extinction values $A_v$ between 0 and 7 with steps of 0.5.}

\begin{figure*}[ht!]
            \epsscale{1.0}
            \plotone{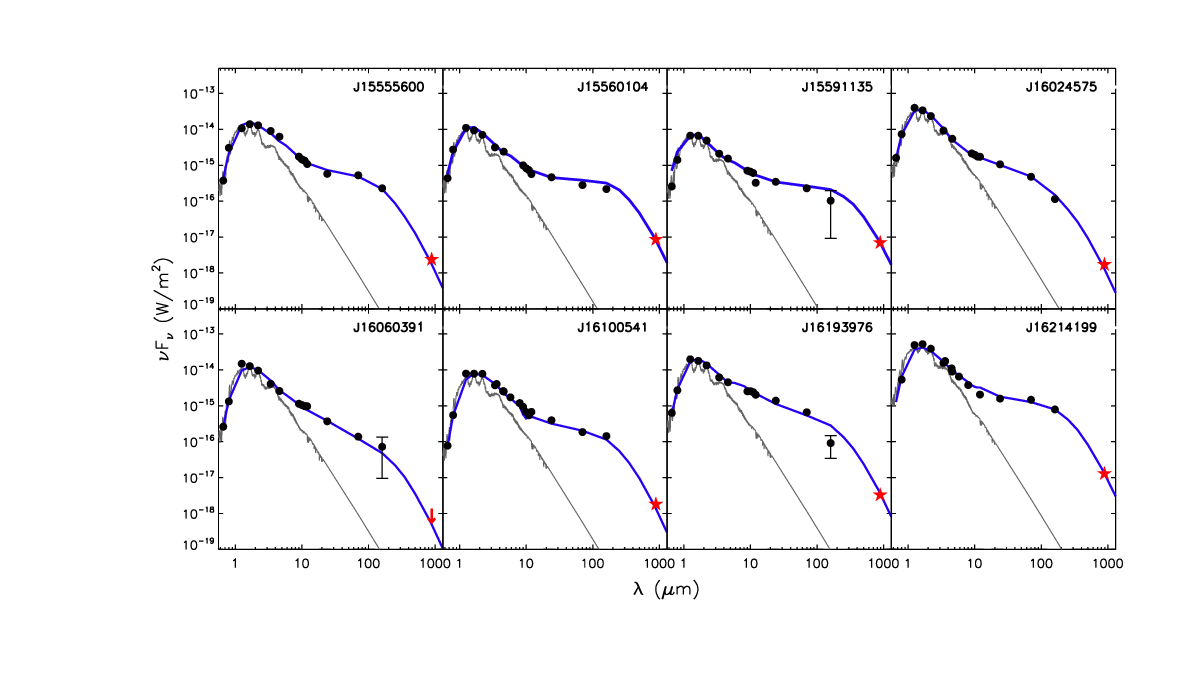}
        \caption{Spectral energy distributions of the sample. The MCFOST best-fit models are shown (blue lines), in addition to representative stellar atmospheric models (grey lines) normalized to 1.6~$\mu$m. Photometry measurements (black points) from optical to far-IR wavelengths are compiled from \citet{2006ApJ...644..364A, 2000AJ....120..479A, 2009AA...504..981B, 2003yCat.2246....0C, 2012yCat.2311....0C, 2012ApJ...744L...1H, 2007ApJ...660.1517S, 2008ApJ...688..377S}. Our ALMA Band 7 (885~$\mu$m) flux densities and upper limit are shown by the red stars and downward arrow, respectively.}
  \label{fig:sedgallery2}
\end{figure*}

\begin{figure*}[ht!]
     \begin{center}
            \epsscale{1.2}
            \plotone{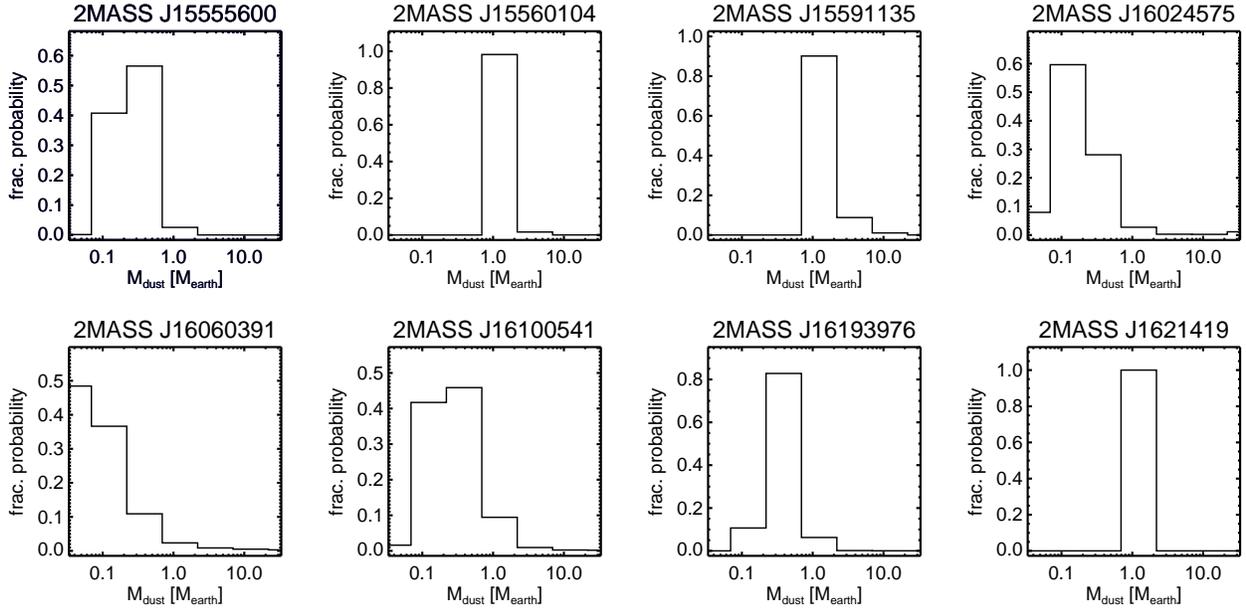}
        \caption{Dust mass probability distributions for all sources within our survey, calculated from the radiative transfer grid presented in \S \ref{sec:analysis}. }
  \label{fig:mcfostdust}
  \end{center}
\end{figure*}

We show the SEDs for the best fit models in Figure \ref{fig:sedgallery2} and the best fit dust masses in Table~\ref{tab:transfer-results} and Figure \ref{fig:mcfostdust}. {\gc For each model, the fitting quality was assessed via the $\chi ^2$ metric. Uncertainties were set to the maximum between the observed value and 10\% of the measured flux, to avoid giving too much weight to optical data points, where variability can be important.  For each model we estimated the Bayesian probability via exp(-$\chi ^2$/2), assuming flat priors. The probability distribution presented in Figure \ref{fig:mcfostdust}, where obtained by marginalising the multi-dimensional probability over each dimension successively.
 We assumed the model uncertainties were negligible compared to the observed ones.} The disk masses we derive here are in agreement with those from our previous study \citep{2012ApJ...755...67H} within a factor of 3, often better. This is a validation of our previous claim that the disk masses are so low for these BDs that their disks are optically thin already in the {\it Herschel} / PACS range (70, 100, and 160 $\mu$m) and that these continuum fluxes are adequate tracers of the disk dust mass, even at 70 $\mu$m. We compare the dust masses derived here with the masses derived by other methods below.

Although the dust masses are well constrained by SED fitting, we caution again that because the disks are optically thin at most wavelengths, SED fitting is not sensitive to the disk geometry and that the disk structural parameters are poorly constrained {\gc, and that uncertainties in the dust opacities directly impact the derived dust masses}.

\begin{table*}
\setlength{\tabcolsep}{3pt} 
\caption{Comparison of disk masses using different approaches to estimate the disk {\gc mass-averaged} dust temperature.} \smallskip
\label{tab:transfer-results}
\scriptsize
\begin{minipage}[t]{\textwidth}
\noindent\begin{tabularx}{\columnwidth}{@{\extracolsep{\stretch{1}}}*{6}{l}@{}}
\hline\hline
Object & Luminosity & Analytic Dust Mass$^{a}$ & Analytic Dust Mass$^{a}$ & Analytic Dust Mass$^{a}$  & Model Dust Mass  \\
     & L$_\odot$ & $T_{d} = 25(L_{*}/L_{\odot})^{0.25}$&  $T_{dust} = 22(L_{*}/L_{\odot})^{0.16}$ & Disk-Averaged from model&  \\
(1) & (2)  & (3) & (4) & (5)  & (6)\\
 
\hline
2MASS J15555600 &  0.0075  & 1.20 (0.88)  & 0.54 (0.39) & 0.13 (0.09) & 0.33  \\
2MASS J15560104 &  0.0075  & 4.35 (3.18)  & 1.96 (1.43) & 1.54 (1.12) & 1.05  \\
2MASS J15591135 &  0.0114  & 3.91 (2.86)  & 1.67 (1.21) & 0.90 (0.65) & 1.05  \\
2MASS J16024575 & 0.0064   & 0.68 (0.49)  & 0.35 (0.25) & 0.21 (0.15) & 0.11  \\
2MASS J16060391 & 0.0045   & 0.32 (0.23)  & 0.12 (0.09) & 0.04 (0.03) & 0.03  \\
2MASS J16100541 & 0.0064  & 1.02 (0.74)  & 0.44 (0.32) & 0.21 (0.15) & 0.33  \\
2MASS J16193976 & 0.0064   & 1.84 (1.34)  & 0.79 (0.57) & 0.43 (0.31) & 0.33  \\
2MASS J16214199 & 0.078    & 1.94 (1.42)  & 1.57 (1.14) & 2.03 (1.47) & 1.05  \\

\hline
\end{tabularx}
\end{minipage}
\tablecomments{ {\gc In column (2) we list the luminosity used for calculating the dust temperatures of the objects in the left column (1). From column (3) onward we list} the masses (M$_\oplus$) calculated following Equation \ref{eq:dustmass} and using temperatures from the \citet{andrews2013} relation (3), an updated version of this relation suitable for BD disks using lower luminosity objects and smaller disk sizes (4), and the {\gc mass-averaged} dust temperatures in out best-fit radiative transfer models (5, both this paper). In the rightmost column (6) we report the disk mass taken directly from the best-fit radiative transfer model for each source.\\
$^{a}$ Parenthetical values indicate scaled masses corresponding to an 885$\mu$m opacity of 3.7 cm$^{2}$/g to match the model parameters. Original values correspond to a scaled 885$\mu$m opacity of 2.7 cm$^{2}$/g (cf. Carpenter et al. 2014).}
\end{table*}

\section{Discussion}
\label{sec:discussion}

In this section we compare our results with previous studies. BDs in Upper Sco are of special interest because they allow us to compare the evolution of disk masses as a function of spectral type in Upper Sco itself, but also as a function of time in comparison with other younger associations. We stress, however, that our sample is by construction significantly biased towards far-IR bright sources, and therefore also biased towards sources with larger disk masses \citep[see][]{2012ApJ...755...67H}.

\subsection{Comparison of dust masses from radiative transfer models and analytical prescriptions}
\label{sec:RTvsAndrews}

An easy way to relate observed sub-mm fluxes to disk masses is by assuming the emitting dust is optically thin and well represented by a single temperature:
\begin{equation}\label{eq:dustmass}
log M_{dust} = log S_{\nu} + 2 log d - log \kappa_{\nu} - log B_{\nu}( \langle T_{dust} \rangle),
\end{equation}
where $S_{\nu}$ is the flux density, $d$ is the distance, $\kappa_{\nu}$ is the dust opacity, and  $B_{\nu}(\langle T_{dust} \rangle)$ is the Planck function evaluated at the {\gc mass-averaged} dust temperature. This equation estimates the dust mass in the disk.
\citet{andrews2013} calibrated the relationship between the {\gc mass-averaged} dust temperature, $\langle T_{dust} \rangle$, and the stellar luminosity, $L_*$, using a suite of radiative transfer models in the luminosity range 0.1L$_{\odot} < L_* < 100L_\odot$. In that range, the dust temperature scales with the stellar luminosity as $\langle T_{dust} \rangle \approx 25 (L_{*} /L_{\odot})^{1/4} K$. 

In Upper Sco, this temperature scaling was used by \citet{2014ApJ...787...42C} to calculate the dust masses around K and early M type stars. Following them, we calculate the dust masses for our targets in a similar way using the same opacity as used by \citet{andrews2013} but scaled to the wavelength of our observations: $\kappa_{885} = 2.7 g^{-1} cm^{2}$ at 885~$\mu$m, assuming $\kappa_{1.3mm} = 2.3 g^{-1} cm^{2}$, $\kappa \propto \nu^{\beta}$, and $\beta$ = 0.4 as in \citet{andrews2013} and \citet{2014ApJ...787...42C}. We call these dust masses the "analytical masses".

\begin{figure*}[tbh]
     \begin{center}
            \epsscale{1.1}
            \plotone{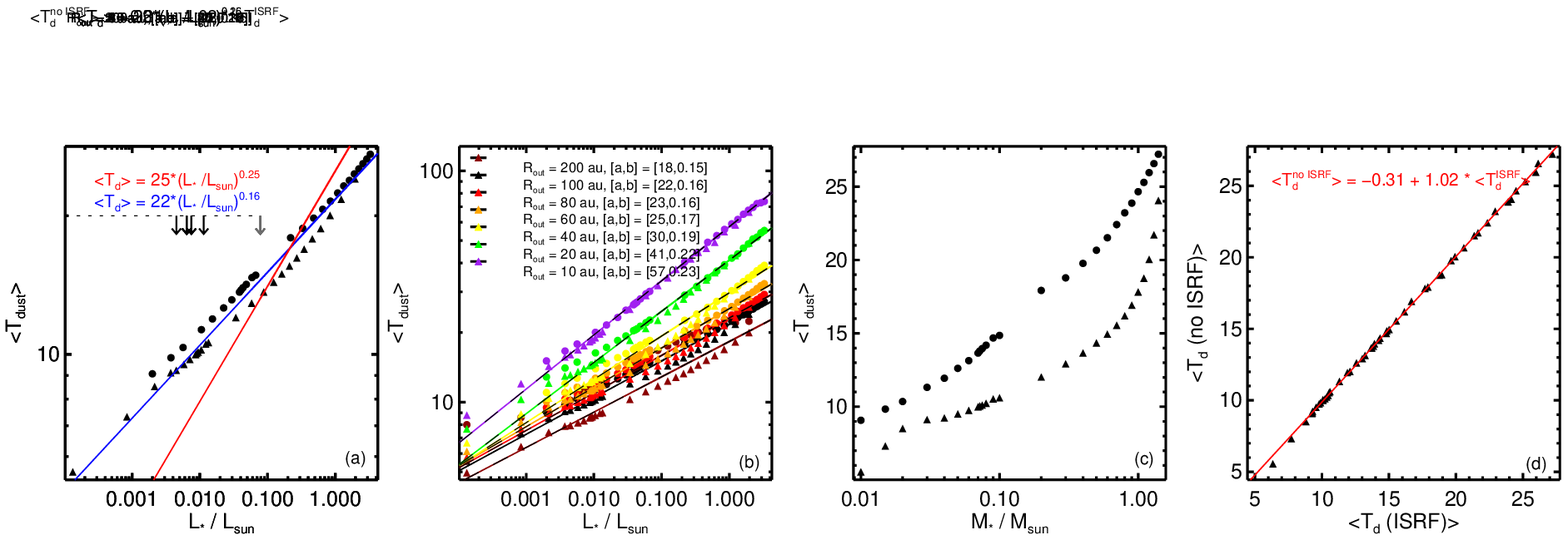}
            \caption{{\gc Mass-averaged model dust temperatures $\langle T_{dust} \rangle$ {\gc for a set of generic disk models disk models} around low luminosity objects with the stellar parameters taken from the 1 {\gc (filled circles)} and 10 Myrs {\gc (filled triangles)} isochrones of \citet{2015A&A...577A..42B}. Shown from left to right: \textbf{(a)} dust temperature as function of central object luminosity, overplotted a powerlaw fit in blue, and the \citet{andrews2013} solution in red. The downward arrows show the luminosities of our Upper Sco (black) and Ophiuchus (grey) objects; \textbf{(b)} the effect of an outer disk radius varying between 10 and 200 au.  Dust temperatures for each outer radius following a powerlaw of the shape $<$T$_d>$ = a~$\times$~(L$_*$ /L$_{\odot}$)$^{b}$ are also shown in the panel; \textbf{(c)} the effect of stellar age on dust temperature; \textbf{(d)} the effect of an external radiation field on dust temperature.}}
  \label{fig:MCFOSTgrid}
  \end{center}
\end{figure*}

The analytical dust masses we derive {\gc under these assumptions} are on average a factor of 3.5 larger than the masses derived using the radiative transfer modeling {\gc described in \S \ref{sec:analysis}}. 
Even if we allow for the difference in dust opacity between the value above and that for our radiative transfer models (3.7 cm$^{2}$/g, see Table~\ref{tab:transfer-results}), the analytical dust masses are still a factor of 2.5 larger than the model masses. We note also that the BDs in our sample have luminosities in the range 0.002L$_{\odot} < L_{BD} < $0.05$L_\odot$, well outside of the range tested by \citet{andrews2013}. This may lead to significant deviations in the calculated disk masses as discussed below. 

One likely candidate for this discrepancy is the very low dust temperature in the disks around late M stars required by the scaling relation. This {\gc low temperature} is motivated by the tight correlation observed between stellar luminosity and \textit{Spitzer} 24~$\mu$m flux \citep{2010ApJS..186..111L, 2010ApJS..186..259R}, which suggests that stellar luminosity provides the sole heating contribution to the isothermal disk regions producing the greatest amount of mm-wave emission \citep{andrews2013}. 

There are several factors that can interfere with this scaling for disks around very low mass objects. One important weakness for example, is that at low luminosities dust heating by the general inter-stellar and intra-cloud radiation field is significant in the outer parts of the disks. 
Another reason why VLMS and BD disk temperatures may deviate from the proposed scaling relation is their possible smaller disk size and/or shape. If the extent of the disk radius is significantly smaller for lower-mass primaries, and/or if the scale height distribution differs, the dust temperature scaling relationship will be accordingly different. If we consider {\gc a} changing disk radius with primary mass as the most significant factor, it has been shown for a benchmark suite of models, including MCFOST and RADMC-3D (used by Andrews et al. 2013), that beyond a saturation point of incident flux at the innermost region of the disk, the dust temperature within the disk falls with radius in a $\ T \sim 1/ {R}^{0.4\ to\ 0.5}$ fashion \citep[Figure 3 of][]{2009A&A...498..967P, 2014A&A...565A.129P}. Finally, as young objects are moving towards the main sequence and contract, their luminosities decrease for similar spectral types. The \citet{andrews2013} grid was constructed using objects of 2.5 Myr, markedly different from the 11 Myr Upper Sco objects.  This in turn has a direct influence on the heating of the disk.

To test the influence of disk size, stellar luminosity {\gc (age)} and external radiation field, we calculate a grid of disks similar to the models described in \S \ref{sec:analysis}, {\gc but with most parameters fixed at a representative value. These fixed values are: a scale height H0 of 10 au,  $\beta$ (the exponent of the scale height profile) = 1.125, an inclination of 45$^{\circ}$, the inner radius R$_{in}$ at 0.1 au, a surface density exponent $p$ of -0.5, and a disk mass fixed at 1\% of the (sub)stellar mass. We vary the disk outer radius R$_{out}$  between 10 and 200 au, and use stellar parameters ($R_{*}$, $M_{*}$ and $T_{*}$) from the \citet{2015A&A...577A..42B} models for objects of 1 and 10 Myr. These models cover a mass range between 0.01 and 1.4 M$_\odot$, and luminosities between 0.00013 and 3.3 L$_\odot$. Results are summarized in Figure \ref{fig:MCFOSTgrid}, For all but the second from the left panel, we show results for disks with an outer radius fixed at 100 au. }

{\gc
We fit a power law to the correlation between luminosity and {\gc mass-averaged} disk dust temperature for the models with an outer radius of 100 au and obtain a best fit represented by $\langle T_{dust} \rangle \approx 22 (L_{*} /L_{\odot})^{0.16} K$.  (left panel Figure \ref{fig:MCFOSTgrid}).  This powerlaw predicts higher values for the disk {\gc mass-average}d dust temperature than the prescription derived for higher luminosity disks \citep{andrews2013}, which is not unreasonable since in their Figure 17 the dust temperature decrease seems to flatten out for lower luminosities. In our parameter space we explore low luminosities largely unexplored by the \citet{andrews2013} study, and we want to stress that the luminosity-temperature scaling relations we derive in this work are valid for spectral types later than $\approx$ M5. Furthermore, SED fitting in absence of spatially resolved data allows finding a good fit to the data up to the point where a small enough optically thick disk cannot reproduce the observed flux anymore. Following Figure 5  from \citet{2008A&A...486..877B}, and keeping into mind the lower fluxes for our objects, this puts a lower size limit on our disks of a few au. Any disk size beyond this lower limit
is possible for our sample, and we remind the reader that as $\langle$T$_{dust}\rangle$ increases with shrinking R$_{out}$, the mass required to fit the SED decreases.

In panel b of Figure \ref{fig:MCFOSTgrid} we examine the dependence of the disk outer radius on the disk {\gc mass-averaged} dust temperature. As expected, the disk outer radius has a large influence on the {\gc mass-averaged} dust temperature, with temperatures as low as 0.8 (r$_{out}$ = 200 au) and as high as 2.8 (r$_{out}$ = 10 au) times the baseline value we set at 100 au. We fit a scaling relation in the form of a $\langle$T$_{dust}\rangle$ = a~$\times$~(L$_*$ /L$_{\odot}$)$^{b}$ for each separate outer radius and note these [a,b] values in the Figure.  
}

The effect of stellar age on the {\gc mass-averaged} dust temperature for stars of equal mass is that $\langle T_{dust} \rangle$ in the disks around the older stars is systematically lower by between 2 K and 7 K, or between 13 and 64\%, compared to $\langle T_{dust} \rangle$ in the disks around the younger (1 Myr) objects (panel c). For the stellar mass range represented by our target list, the average dust temperature in the 10 Myr old grid is between 30 and 50\% lower compared to the 1 Myr old grid. This effect is, of course, directly driven by the decreasing luminosity as both the temperature and stellar radius decrease with age.

Finally, we tested for the influence of an external {\gc (interstellar)} radiation field {\gc (ISRF)} by including a cosmological background + diluted O-B stars field, described in \citet{2009A&A...501..383W}. The effect of the contribution from an external radiation field on the disk temperature is at most a few percent (right panel of Figure \ref{fig:MCFOSTgrid}) and is negligible for our purposes.


\subsection{Comparison with objects of earlier spectral types in Upper Sco.}\label{sec:dust-disk-usco-allmass}

\begin{figure}[ht!]
            \epsscale{1.2}
            \plotone{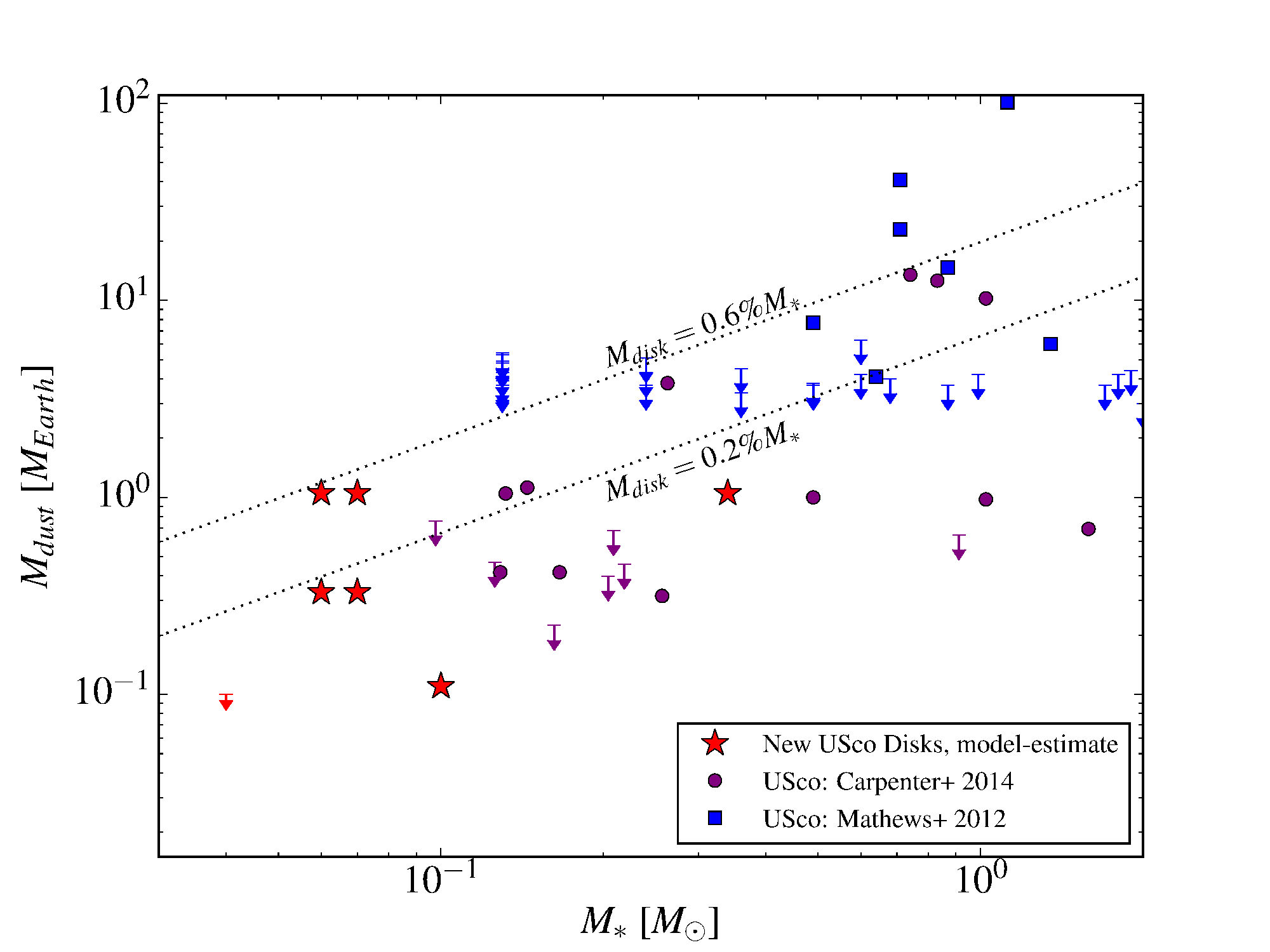}
        \caption{Derived disk dust mass as a function of central object mass. Shown are the new Upper Sco targets from this survey (star symbols), with dust masses calculated from the 885 $\mu$m continuum flux. Shown for comparison are a recent survey of higher primary mass Upper Sco members from  \citet[][purple circles]{2014ApJ...787...42C} and \citet[][blue squares]{mathews2012}. Overplotted with dotted lines are lines of constant disk mass to stellar mass ratios of 0.2\% and 0.6\% assuming a dust-to-gas ratio of 100. 3$\sigma$ upper limits are denoted by downward arrows.}
\label{fig:dustmass}
\end{figure}

The largest study of disk evolution for Upper Sco is that of \citet{luhman2012}. Based on {\it Spitzer} and {\it WISE} data, both sensitive to emission from the inner disk but not to dust mass, they conclude that disks are longer lived around the lower mass members of the association than around the earliest spectral types.  Since total disk mass cannot be traced using wavelengths where the emission is optically thick, this does not directly address the evolution of the total mass of the disks.  Millimeter and sub-mm observations of Upper Sco disks \citep{mathews2012, 2014ApJ...787...42C} including our own have observed a modest number of total objects, but there does appear to be a trend of disk mass with host mass.  Our observations extend the host mass down to a $few \times 10^{-2}$ M$_\odot$.  As shown in Figure \ref{fig:dustmass} for all the mm/sub-mm observations including ours, there is some correlation of disk mass and host mass with, albeit, substantial scatter.  If we view our data as sampling the upper end of the distribution of disk masses for Upper Sco BD/VLMS, then this rough correlation is still present in the comparison with earlier type young objects in Upper Sco.

\subsection{Declining detection rates for the gas disks with time}

We detect $^{12}$CO J=3-2 emission in 1 of the 7 disks in Upper Sco. This disk, around the M6.5 object 2MASS J15555600, has an estimated dust mass of 0.33 M$_\oplus$, which is one of the weaker detections in our sample. An emission line with similar line/continuum brightness would have been detected around any of our other sources. 
Yet, dust should become optically thinner than CO very rapidly if the mass is reduced similarly for dust and gas. Keeping in mind that our sample is selected to cover the brightest disks in this spectral type range in Upper Sco, these low detection statistics suggest that most of the gas-phase CO in these old disks has decreased or disappeared, either by freeze out onto dust grains, accretion onto the star, or photo-evaporation. While we still detect dust emission, it is also possible that these disks are transitioning into debris disks

We estimate a lower limit to the emitting gas mass of 0.07 - 0.43 M$_\oplus$ by following the prescription from \citet{2015arXiv150904589R} and references therein, adopting a CO:H ratio of 10$^{-4}$, a line optical depth of $\tau$ = 1 and a temperature range of 7.5 - 26.6 K (see Table \ref{tab:transfer-results}), equal to the {\gc mass-averaged} dust temperatures found for 2MASS J15555600. Note that this number ignores the frozen out CO reservoir, and that $^{12}$CO J=3-2 emission in general has optical depths exceeding 1. Changing the optical depth modifies the CO mass with a factor $\tau$ / (1 - e$^{-\tau}$).

One possible explanation as to why we detect CO emission in the disk around 2MASS J15555600 but not around any of the other disks in Upper Sco is that this disk has a H-[4.5] photometric color of 2.17, while that value for the rest of our sources is between 1.07 and 1.81. This color is a proxy for the warm dust in the inner disk, which suggests that the inner dust disk of 2MASS J15555600 is warmer compared to the other targets. This in turn could boost the emissivity of the CO gas. Interestingly, this photometric color is similar to the color of 2 other VLMS / BD disk found in Taurus by \citet{ricci2014} and detected in CO, 2M044 (2.36, spectral type M7.25) and CIDA 1 (2.45, spectral type M5.5). 

\section{Conclusions}

We summarize our conclusions as follows:
\begin{enumerate}
\item We detect 885 $\mu$m emission from 6 out of 7 surveyed BD disks in the Upper Sco star forming region, and from one disk in Ophiuchus.
\item We find that the dust masses of the detected disks are low, between 0.1 and 1 M$_\oplus$. The ratio of these disk masses to the central object mass is in agreement with previous results of \citet{mathews2012} and \citet{andrews2013} on the disk / central object mass ratio for disks around stellar mass objects, and suggests that this correlation can be extrapolated down to the BD regime.
\item The correlation between stellar luminosity and {\gc mass-averaged} dust temperature for BD disks can be approximated by $\langle T_{dust} \rangle \approx 22 (L_{*} /L_{\odot})^{0.16} $ K {\gc for disks with an outer radius of 100 au, but increases with shrinking disk outer radius. See Figure \ref{fig:MCFOSTgrid} for scaling relations for other outer disk radii.}
\item We detect $^{12}$CO J=3-2 emission in two disks, one of which belongs to Upper Sco: 2MASS J15555600 (spectral type: M6.5). The other detection is in the disk around 2MASS J16214199 in Ophiuchus.
\item The disk of 2MASS J15555600 appears warm compared to other Upper Sco targets in our sample based on the H-[4.5] photometric color. Interestingly, this photometric color is similar to the color of two other VLMS / BD disks with detected CO emission found in Taurus by \citet{ricci2014}.
\end{enumerate}

In this manuscript we present the first part of our survey of disks around stars crossing the stellar/substellar boundary. We will present a similar analysis on disks in the Taurus star forming region in Ward-Duong et al. (in prep, for the continuum emission) and in van der Plas et al. (in prep, for the CO emission). 

\acknowledgments

Gvdp and SC acknowledge support from the Millennium Science Initiative (Chilean Ministry of Economy) through grant RC130007. GP acknowledges financial support from FONDECYT, grant 3140393 and SC acknowledges support from FONDECYT grant 1130949. KWD is supported by a National Science Foundation Graduate Research Fellowship under Grant No. DGE-1311230 and a NSF Graduate Research Opportunities Worldwide supplemental award (Proposal \#13074525), in partnership with CONICYT. PH was partly founded by the Joint ALMA Observatory Visitor Programme. This paper makes use of data from ALMA programme 2012.1.00743.S. ALMA is a partnership of ESO (representing its member states), NSF (USA) and NINS (Japan), together with NRC (Canada) and NSC and ASIAA (Taiwan), in cooperation with the Republic of Chile. The Joint ALMA Observatory is operated by ESO, AUI/NRAO and NAOJ. The National Radio Astronomy Observatory is a facility of the National Science Foundation operated under cooperative agreement by Associated Universities, Inc. {\gc We thank the referee for the helpful suggestions which have strengthened and clarified this manuscript.}


Facilities: \facility{ALMA}

\end{document}